\documentclass[conference]{IEEEtran}

\usepackage{booktabs} 

\usepackage{hyperref}
\usepackage{graphicx}
\usepackage{mdframed}
\mdfdefinestyle{style1}{leftmargin=1cm,rightmargin=0.5cm,skipbelow=0.5cm,skipabove=0.2cm,
   innertopmargin=2pt}
	\usepackage{amsmath}
	\usepackage{amsfonts}
    \usepackage{longtable}
	\usepackage{fancyhdr}
	\usepackage{mdframed}
\mdfdefinestyle{style1}{leftmargin=1cm,rightmargin=0.5cm,skipbelow=0.5cm,skipabove=0.2cm,
   innertopmargin=2pt}
	\usepackage{balance}
    \usepackage{fancybox}
    \usepackage{booktabs}
	\usepackage{enumerate}
    \usepackage{framed}
    \usepackage{pgfplots}
	\usepackage{listings}
    \usepackage{subcaption}
    \usepackage{xcolor}
    \usepackage{hyperref}
  \DeclareGraphicsExtensions{.pdf,.png}
  
\ifCLASSOPTIONcompsoc
  \usepackage[nocompress]{cite}
\else
  \usepackage{cite}
\fi

\begin{document}
\title{Agent-Based Software Testing: A Definition and Systematic Mapping Study}


\author{Pavithra Perumal Kumaresen$^{1}$, Mirgita Frasheri$^{1}$, Eduard Paul Enoiu$^{1}$
\\$^{1}$M\"alardalen University, V\"aster\aa s, Sweden%
}

\IEEEtitleabstractindextext{%
\begin{abstract}
The emergence of new technologies in software testing has increased the automation and flexibility of the testing process. In this context, the adoption of agents in software testing remains an active research area in which various agent methodologies, architectures, and tools are employed to improve different test problems. Even though research that investigates agents in software testing has been growing, these agent-based techniques should be considered in a broader perspective. 
In order to provide a comprehensive overview of this research area, which we define as 
agent-based software testing (ABST), a systematic mapping study has been conducted.
This mapping study aims to identify the topics studied within ABST, as well as examine the adopted research methodologies, identify the gaps in the current research and point to directions for future ABST research.
Our results suggest that there is an interest in ABST after 1999 that resulted in the development of solutions using reactive, BDI, deliberative and cooperate agent architectures for software testing. In addition, most of the ABST approaches are designed using the JADE framework, have targeted the Java programming language, and are used at system-level testing for functional, non-functional and white-box testing. In regards to regression testing, our results indicate a research gap that could be addressed in future studies.


\end{abstract}
}
%
%



\maketitle
\IEEEdisplaynontitleabstractindextext
\IEEEpeerreviewmaketitle

\section{Introduction}
Nowadays, continuous development and integration processes are subject to large and frequent changes. Consequently, software development organizations need to deliver
reliable and high-quality software products while having to consider more stringent
time constraints. A side effect of such constraints is the limitation in the amount of development and testing that can be performed before delivering the software~\cite{ammann2016introduction}. Intelligent and automated techniques can be used to tackle this problem. For example, agents have already been used to automate different aspects of testing and improve test efficiency and effectiveness~\cite{decker1997intelligent,malz2012prioritization}. 

Agents are software systems that operate in an environment which they can perceive and act upon, while also being able of autonomous actions~\cite{wooldridge1997agent}. Depending on the flexibility of such actions, agents might be able to take initiatives and select their own goals, and interact with others when deemed fit. What is missing from the state of the art is a comprehensive approach for defining agent-based software testing and its applications.

To tackle this gap, first we propose a definition of the research area as Agent-Based Software Testing (ABST). Secondly, we present the results from a systematic mapping study, in which we identify the areas of application together with the tools, techniques, and methods used in the development of agent-based systems for software testing, while also examining the research methodologies adopted in these different works. 
Based on the results obtained from this study, we identify research trends and gaps which can be useful for both researchers and practitioners. 

\section{A Definition of Agent-Based Software Testing}

Agents are software systems -- which could be embodied into physical entities -- that operate in an environment which they can perceive and act upon, and are able of performing autonomous actions~\cite{wooldridge1995intelligent}. Agents are used in many domains, can take different physical forms~\cite{balaji2010introduction}, and have different properties such as: (i) operating on their own without human interventions (autonomy), (ii) interacting with other agents (social ability), (iii) perceiving their environment and responding in a timely fashion to changes that occur in it (reactivity), and (iv) taking initiative to exhibit a goal-oriented behavior (pro-activeness). An agent can be capable of learning, by acquiring new knowledge and skills, which can be used to take better decisions in the future. 

Agents have already been used to automate different aspects of software development~\cite{wooldridge1997agent, winikoff2009future,erlenhov2019current}.  Several researchers have proposed different approaches for using agents specifically in software testing~\cite{tang2010towards,salima2007enhancing,dhavachelvan2005multi}, by considering different aspects that relate to test management, test design, execution and evaluation. Jeff Offutt~\cite{wotawa2016testing} outlined in a keynote from 2016 that there is a need in test automation for more intelligent tests exhibiting self-determination and self-awareness. However, the relation and overlap between the use of (intelligent) agents in software testing and what part of the testing process these methods optimize is unclear. Hence, we propose that \textit{Agent-Based Software Testing (ABST) should be defined as the application of agents (e.g., software agents, intelligent agents, autonomous agents, multi-agent systems) to software testing problems by tackling and automating complex testing tasks}. 

The aim of ABST research is to address software testing problems through agent-based paradigms, by additionally using the variety of techniques from artificial intelligence and software engineering. We realize that the process of defining the ABST research area is evolutionary and iterative. The ABST definition needs to be further discussed as the evidence and knowledge in this area grows and refines. 


 \begin{figure*}[tbp]
	\centering
    \includegraphics[width=0.9\textwidth]					{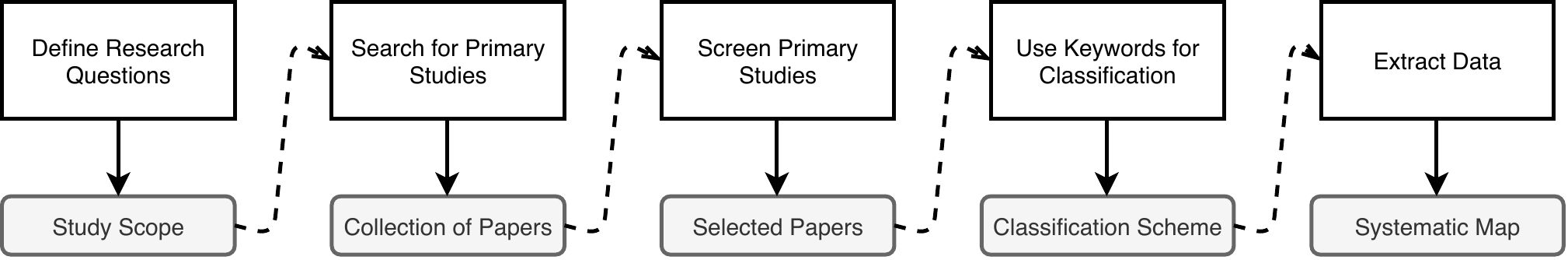}  
    \caption{The Mapping Study Steps Performed in this Study.}\label{fig:3}
\end{figure*}

\section{Research Method}

A systematic mapping involves several steps like identification of papers, analysis, and classification of selected papers in the area of interest (i.e., agent-based software testing). The systematic mapping study is performed in accordance with the steps shown in Figure~\ref{fig:3} based on the guidelines of Petersen \textit{et al.}~\cite{petersen2008systematic}. There are five steps involved in performing this mapping study starting with the definition of the research questions which gives a scope for the search of studies. Based on the scope of the research, a set of search strings are derived and applied in the selected databases to identify studies within this research area. From the identified studies, the most relevant ones are screened by applying certain inclusion and exclusion criteria. We extracted the abstract descriptions, introduction, and conclusion sections of the filtered papers and we studied these to identify the keywords used during the classification stage for answering our research questions. Finally, these papers are studied in-depth to extract the data for each category dimensions and then the obtained results are discussed to show the final systematic map. 

\subsection{Definition of the research questions}

The main objective of this section is to devise a set of research questions (RQs) based on the definition of ABST and better frame the scope of this research. 
To this end, we formulate three research questions. The main objective of the first question is to identify the research interest and types of contributions in this domain:

 \begin{mdframed}[style=style1]
 {\it RQ1: What is the current state of agent-based software testing research?}
 \end{mdframed}
 
As RQ1 is a broad research question, thus four sub-questions (RQ1.1– RQ1.4) have been identified:
\begin{itemize}
\item RQ1.1: What number of academic studies on agent-based software testing has been published?
\item RQ1.2: What are the publication channels used to publish studies on agent-based software testing?
\item RQ1.3: What kinds of contributions are provided by studies on agent-based software testing?
\item RQ1.4: What research methods have been used in empirical studies on agent-based software testing?
\end{itemize}

The main objective of the second research question is to identify the method-related characteristics of the agent-based approaches:

 \begin{mdframed}[style=style1]
 {\it RQ2: What are the characteristics of these agent-based systems used in software testing?}
 \end{mdframed}

This question is divided into the following three sub-questions focusing on agent-based systems' architecture and implementation:
\begin{itemize}
\item RQ2.1: What are the agent architectures used in agent-based software testing?
\item RQ2.2: What are the development frameworks used for implementing agent-based software testing?
\item RQ2.3: What are the programming languages adopted by agent-based software testing?
\end{itemize}

The main objective of the third question concerns test-related characteristics of agent-based systems used for software testing:

 \begin{mdframed}[style=style1]
 {\it RQ3: What are the testing characteristics provided by the agent-based software testing approaches?}
 \end{mdframed}
This question was divided into three sub questions as follows:

\begin{itemize}
\item RQ3.1: What are the testing levels targeted by the agent-based software testing approaches?
\item RQ3.2: What are the testing areas targeted by agent-based software testing?
\item RQ3.3: What are the types of applications targeted by these agent-based software testing approaches?
\end{itemize}

\subsection{Search process}
Searching for papers is a critical phase in a systematic mapping study as it ensures the comprehensive coverage of the research topic under study. We devised a set of search strings for conducting the search in multiple digital libraries in the ABST research area. Based on the ABST definition we used the following search string: "Agent AND Software AND Testing". The search format varies with different databases as follows:
\begin{itemize}

    \item IEEE: (("Abstract”: software) AND ("Abstract”: testing) AND ("Abstract”: agent)) 
    \item ACM: [Abstract: software] AND [Abstract: testing] AND [Abstract: agent] 
    \item SCOPUS: TITLE-ABS-KEY (“Software” and “testing” and “agent”)

\end{itemize}

The following databases were selected to perform the search: IEEE Xplore digital library, ACM digital library and the SCOPUS scientific database. The selected databases are pertinent to this study as these return the most manually collected publications on ABST. The search string used across these databases retrieved 2663 papers (i.e., 667 papers for IEEE, 860 papers for ACM, and 1136 for Scopus).

We mention here that there are other methods that can be used to improve the coverage of such a mapping study. One example of such a method is the snowballing (backward and forward) search~\cite{wohlin2014guidelines}. The selection of a start set of papers used to perform snowballing search is one of the main challenges in this procedure. Snowballing is not necessarily an alternative to the database search process. Due to the nature of this study, we used a different approach to check the reliability of the relevant literature and ensure the best possible coverage of the literature. We used the comprehensive related work collected in a paper outlying the area of test agents by Enoiu and Frasheri~\cite{enoiu2019test}. We selected these papers and checked their inclusion against the results obtained from our search process. All papers included in this paper have been found using our search strategy and databases. However, a detailed snowballing process could potentially extend the confidence in our search process.

\subsection{Paper Screening}
The process of screening was performed to refine the search results by eliminating the duplicated entries and the non-related ABST papers based on certain screening criteria. A set of inclusion and exclusion criteria were applied to the title, abstract, and keywords sections of the papers to identify the relevant ones. If the data from these sections was insufficient to obtain a decision, we used the introduction and conclusion sections to apply the criteria (as shown in Table~\ref{paperscreening}). Out of the 2663 papers obtained from the search results a set of 48 papers were selected. These 48 papers were studied in detail to perform the data extraction. Due to lack of relevant data with respect to the research questions seven papers were excluded. Finally a set of 41 papers was obtained. The final set of primary studies is available also as a data set~\cite{pavithra_perumal_kumaresen_2020_3874307}.         
\begin{table} [t!]
    \small
    \centering
    \begin{tabular}{|l||l|l|l|}
    \hline
   Databases & Initial Search & Screening & Detailed Study \\
     \hline
    IEEE&	667&	26&	22\\
      \hline
ACM&	860	&9	&7\\
  \hline
SCOPUS	&1136&	13&	12\\
        \hline
    \end{tabular}
    \caption{Paper Screening.}
    \label{paperscreening}
\end{table}

In the inclusion criteria we focused on the relevant aspects by including: (1) papers that are relevant to using agent-based systems in software testing and (2) papers that include the development and maintenance of agent-based systems used in software testing. In addition we used the following exclusion criteria for: (1) papers that are not relevant to agent-based systems, (2) papers on software testing using search-based techniques using machine learning, (3) papers that are not available in full text, (4) papers on the usage of agents to maintain and increase the system performance or other criteria rather than testing the software systems, and (5) papers on testing agent-based systems rather than the usage of these systems in software testing.

\subsection{Classification}
The classification scheme is composed of the following facets based on our research questions: publication and research trends, agent characteristics and testing characteristics. The results obtained in this study are based on manual data extraction and analysis which is done iteratively and revised continuously to mitigate the risk of missing relevant data. The classification scheme used for data extraction is developed based on the revised keywords and adaptations of the commonly accepted and used categorizations. These categories are well established in both the agent and testing areas of research.

\subsubsection{Publication and Research Trends}
We used the following categories for extracting data from each paper in relation to RQ1:
\begin{itemize}
    \item (RQ1.1) Publication rate. It identifies the number of publications in ABST area within a specified period of time. The period of time is defined from the publication dates of the included publications.
    \item (RQ1.2) Publication type. It identifies the channels of publications in ABST area (e.g., journal papers, conference papers, workshop papers).
    \item (RQ 1.3) Research Contributions. An existing classification of research approaches by Wieringa et al.~\cite{wieringa2006requirements,petersen2008systematic} was used in this mapping study: validation research, evaluation research, solution proposal, philosophical papers, opinion papers and experience papers.
      \item (RQ 1.4) Empirical Research Methods. For the empirical studies concerned with ABST we identified three different study types: experiment, case study or comparative study.
\end{itemize}

\subsubsection{Agent characteristics}
This category covers the characteristics of the agent-based systems (RQ2) used for software testing. It consists of the following sub-facets corresponding to each sub-research question:
\begin{itemize}

    \item (RQ 2.1) Agent Architecture. To have an overview of the system structure, it is important to understand its architecture, showing how the parts of an agent system interact~\cite{girardi2013survey} with each other. The agent architecture is considered as the functional controller of an agent involved in making decisions and reasoning to solve problems and achieving goals~\cite{chin2014agent}. Over the years, several surveys and reviews \cite{girardi2013survey,chin2014agent} have been conducted on agent architectures. The agent architecture classification used in this study was based on the classification scheme proposed by Friedenberg and Silverman~\cite{friedenberg2011cognitive}: 
\begin{itemize}
\item	Simple Reactive. Agents in this kind of system are based on response and reasoning facilities similar to that of logical sequences. Nevertheless, in unpredictable and dynamic circumstances they lack decision making capabilities and rely on other agents.
\item	Reactive/Subsumptive. In this architecture, agents are viewed as a collection of simple behavior modules that have a hierarchical organization.
\item	Belief-Desire-Intention (BDI). In this architecture, agents can store the state of their system and environment (beliefs), maintain goals (desires), and contain means to convert the belief and desires into actions (intentions).
\item	Deliberative. Agents have the potential to solve complex problems, including provision for planning, and can perform a sequence of actions to achieve a certain goal. These agents take advantage of contemporary AI technologies (e.g., neural nets, Fuzzy Logic).
\item	Blackboard/Co-Operative. These agents act as a team, where they obtain the state knowledge of the system through a central common agent (blackboard).
\end{itemize}

\item (RQ 2.2) Development Frameworks and Tools. This category identifies the development tools used for each ABST approach. An agent-based system can be designed and built with a help of a toolkit or a framework. Several reviews have been conducted to identify the tools that are used for developing intelligent agent technologies~\cite{rendon2006review,leon2015review}. JADE, JADEX, JACK and MADKIT are examples of agent development frameworks. This category aims to identify the development frameworks used for implementing ABST approaches.

\item (RQ 2.3) Programming Language. This category identifies the commonly used programming languages targeted by agent-based systems~\footnote{Some of the popular programming languages are Java, Python, C, Ruby, and C\#. Data according to L. Kim, "10 Most Popular Programming Languages Today, Available: http://www.inc.com/larry-kim/10-Most-Popular-Programming-Languages-Today.html. [Accessed 11 April 2020].}.
\end{itemize}

\subsubsection{Testing Characteristics}
This category covers the following test-related characteristics of the agent-based systems (RQ3) used for software testing:
\begin{itemize}
   \item (RQ 3.1) Testing Level. This category classified the papers based on the level of abstraction at which ABST is performed. For our purpose we used the following testing levels by adapting some existing classifications \cite{ammann2016introduction,spillner2014software}:
   \begin{itemize}
       \item Acceptance Testing. It assesses the software with respect to its customer requirements.
       \item System Testing. It assesses the software with respect to its architectural design.
       \item Integration Testing. It checks the proper integration of lower units and the correct operation.
       \item Unit/Component Testing. It assesses software with respect to its implementation.
   \end{itemize}

   \item (RQ 3.2) Testing Areas. The following classification of testing areas is based on the categorization proposed by Spillner et al. \cite{spillner2014software}:
\begin{itemize}
  \item Functional Testing. Test cases are created based on the functional requirements of the system. 
    \item Non-Functional Testing. Test cases are designed based on attributes describing the system as a whole and some of its non-functional characteristics such as reliability, usability and performance.
    \item White-Box/Structural Testing. It uses the internal structure of the system for creating test cases.
    \item Regression Testing. The purpose of regression testing is to check whether changes to existing software have introduced errors to functionality that performed correctly in the software's previous version.

\end{itemize}
   \item (RQ 3.3) Types of Applications Under Test. This category identifies the type of applications that are tested using agent-based systems. Agents are used in a wide range of application domains. We extracted data from each paper related to the application under test. The identified types are: web based applications, network applications, industrial applications, distributed software applications.

\end{itemize}

 \begin{figure}[t!]
	\centering
    \includegraphics[width=0.48\textwidth]					{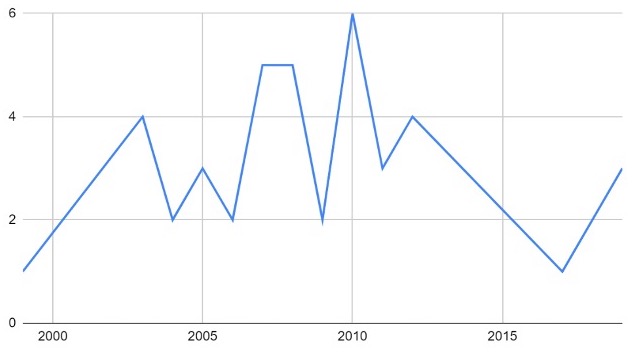}  
    \caption{ABST publications by year.}\label{fig:trends}
\end{figure}

Thus, a simple classification scheme has been formed based on the research questions, which allows for a direct extraction of data items to each facet and reduces the complications of further deeper analysis for the comparison of the results.

\subsection{Data extraction and Mapping}
A final set of papers was obtained by applying the inclusion and exclusion criteria. These papers were studied in detail to extract the data under each category of the classification scheme using an excel sheet. Thus, for every paper that was studied, the respective column details were entered for each facet. Once the data items were identified, a systematic map was developed.

\section{Data Analysis and Results}
The results of this mapping study were analyzed both quantitatively and qualitatively. A quantitative analysis gives the quantitative results under each category whereas the qualitative analysis is used to extract information based on interpretation. 

\subsection{Publication and Research Trends}
The aim of this research question is to establish the annual number of academic studies on ABST as well as the main channels where ABST approaches are disseminated.


\subsubsection{RQ 1.1}  Figure~\ref{fig:trends} lists the number of publications by year of the primary studies from 1999 until 2020. This categorisation is valuable as it indicates that although the number of academic studies on ABST remains rather low, there is a continuous interest over the last decade (peaking in 2010).

\subsubsection{RQ 1.2} The aim of this research question is to identify the main channels where ABST studies are disseminated. The results suggest that 78\% of the primary papers were published in peer-reviewed international conferences, 17\% in journals and just 5\% were published in workshops. 

\subsubsection{RQ 1.3} The classification results based on the research contributions show that 50\% (21 papers) of the studies proposed an ABST solution, while nearly 40\% (16 papers) have contributed by validating ABST approaches and only 10\% (4 papers) have evaluated their proposed ABST solutions. These results highlight a need for ABST research to provide more significant practical contributions, as well as widening the proposed academic solutions towards practical evaluations.


\begin{table} [t!]
    \small
    \centering
    \begin{tabular}{|l||p{6cm}|}
    \hline
   Architecture & Primary Studies \\
     \hline
Reactive &	\cite{lanslots2003using}, \cite{el2006multi}, \cite{zhang2008mobile}, \cite{friess2009multi}, \cite{manzoor2011autonomous}, \cite{chen2012kind}, \cite{tang2010towards}, \cite{salima2007enhancing}, \cite{grundy2005deployed}, \cite{dhavachelvan2006new}, \cite{dhavachelvan2005multi}, \cite{devasena2012multi}, \cite{zhu2004cooperative}, \cite{narendra2005large}, \cite{gardikiotis2007employing}, \cite{da2010jaaf+} \\
     \hline
BDI	& \cite{miclea2003agent}, \cite{kung2004agent}, \cite{ma2010design}, \cite{zhao2010research}, \cite{nguyen2008ecat}, \cite{bai2011design}, \cite{winikoff2017bdi}, \cite{enyedi2008agent} \\
     \hline
Co-Operative &	\cite{huo2003multi}, \cite{liu2003agent}, \cite{chengqing2007ipv6}, \cite{gao2009automatic}, \cite{jordan2019framework}, \cite{enoiu2019test}, \cite{maamri2007maest}, \cite{farias2012distributed} \\
     \hline
Deliberative &	\cite{mala2007intelligentester}, \cite{malz2010agent}, \cite{junior2010improving}, \cite{malz2011agent}, \cite{malz2012prioritization}, \cite{ostrowski1999knowledge}, \cite{mala2008intelligentester}, \cite{karlsson2019exploratory} \\
     \hline
    \end{tabular}
    \caption{Agent Architectures used in ABST research.}
    \label{agentarchitecture}
\end{table}

\begin{table*} [t]
    \small
    \centering
    \begin{tabular}{|l||p{11cm}|p{3.5cm}|}
    \hline
   Frameworks & Description & Primary Studies \\
     \hline
ADK	& ADK (Agent Development Kit) is a commercial agent platform that mainly emphasizes on the mobility and security aspects. &	\cite{miclea2003agent} \\
     \hline
AGLET &	A java mobile agent platform.	& \cite{zhang2008mobile} \\
     \hline
FIPA OS & An open-source FIPA compliant software framework using a simple task-based approach as an internal agent structure.	& \cite{lanslots2003using} \\
     \hline
JACK &	A framework for multi-agent system development for BDI software.&	\cite{mala2008intelligentester} \\
     \hline
JADE &	JADE (Java Agent DEvelopment Framework) is a framework used to build multi-agent systems implemented in Java programming language and complies with the FIPA specifications. & \cite{gao2009automatic}, \cite{ma2010design}, \cite{zhao2010research}, \cite{manzoor2011autonomous}, \cite{malz2011agent}, \cite{malz2012prioritization}, \cite{salima2007enhancing}, \cite{devasena2012multi}, \cite{da2010jaaf+}, \cite{bai2011design} \\
     \hline
JADEX &	A JADE extension developed for rational agents, the BDI model and focusing on web services. &	\cite{nguyen2008ecat} \\
     \hline
JADE LEAP &	JADE-LEAP is a modified version of the JADE platform that can be used in any devices which are java enabled. &	\cite{miclea2003agent}, \cite{mala2007intelligentester}
\\
     \hline
    \end{tabular}
    \caption{Agent Development Frameworks and Tools.}
    \label{tools}
\end{table*}

   \begin{figure*}[t!]
	\centering
    \includegraphics[width=0.9\textwidth]					{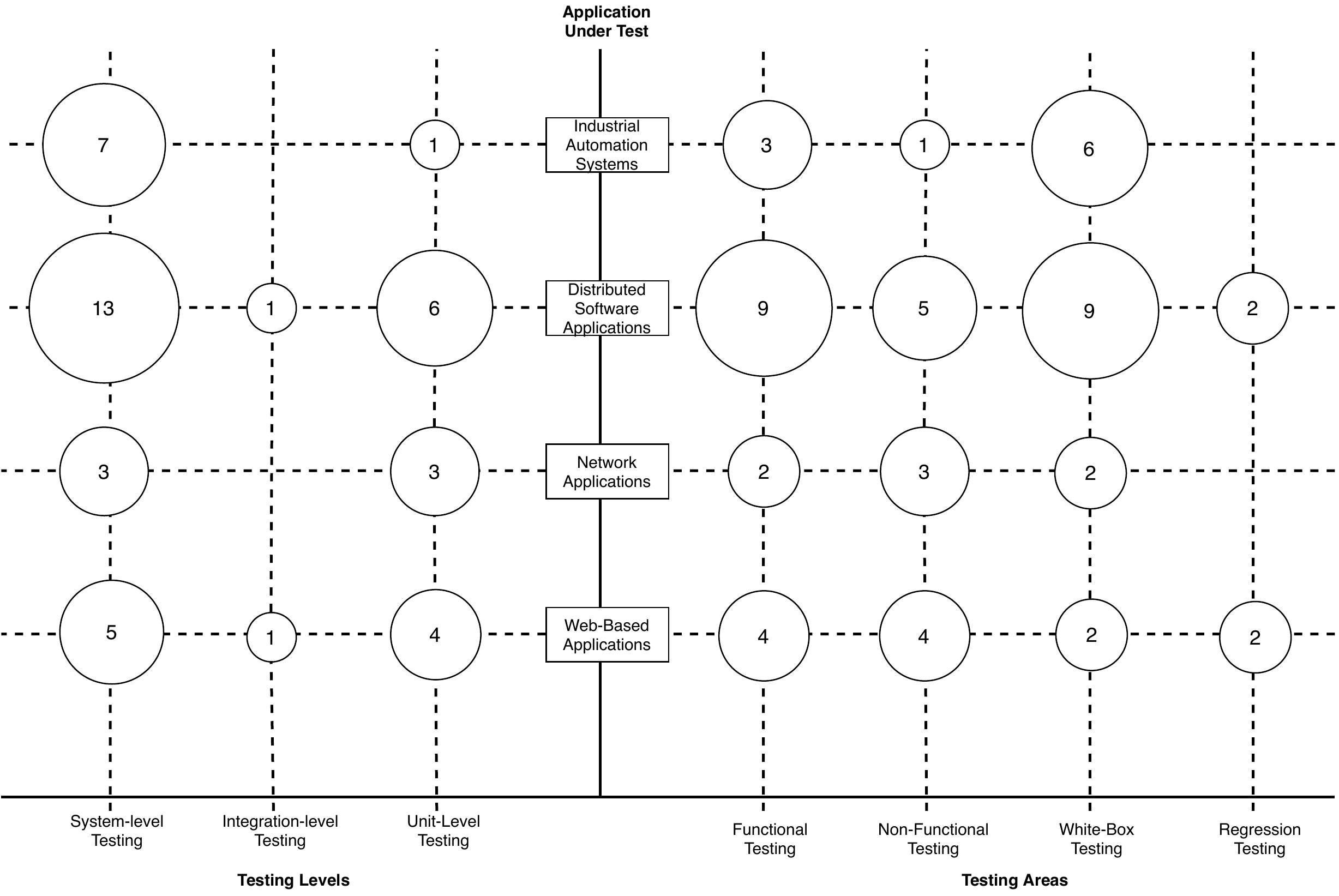}  
    \caption{Distribution of ABST research over a range of testing characteristics, testing levels and applications.}\label{fig:levWels}
\end{figure*}

\subsubsection{RQ 1.4} The analysis of the results related to the empirical research methods used in ABST research shows that 23 studies have conducted experiments in this research area, 4 studies used the case study method while 2 papers perform a comparative study. On the other hand, 32\% of the studies did not use any empirical research methods.

 \begin{mdframed}[style=style1]
 {\it Answer RQ1: There is a continuous interest in ABST after 1999 with the majority of studies being disseminated in conferences and proposing both solutions and performing empirical studies.}
 \end{mdframed}

\subsection{Agent Characteristics}
The analysis under this category gives an overall understanding of the agent-based system characteristics used in ABST.

\subsubsection{RQ 2.1}
The aim of this research question is to categorise studies on ABST based on key agent architectures emerging from the papers being studied. We show the overall results in Table~\ref{agentarchitecture}. 39\% of the primary studies are categorized as using the reactive architecture, 60\% (20\% for each of the other categories) of the studies used BDI, deliberative and co-operative architectures. For one study \cite{mirgorodskiy2008diagnosing} we could not categorize the ABST approach based on the information provided on the used agent
architecture.


\subsubsection{RQ 2.2} The aim of this research question is to identify the reported frameworks used for developing ABST solutions. In Table~\ref{tools} we show seven agent-development frameworks used in ABST. In 61\% of the studies, the ABST solutions are proposed as methods and not implemented in practice. Our results suggest that the most common platform used for ABST development is JADE (10 papers). Some other platforms used are ADK, IBM Aglet, JADE LEAP and JADEX (extensions of JADE), FIPA OS and JACK Intelligent Agent. The scarcity of ABST solutions being implemented would indicate that this research is currently rather restricted and has not yet scaled to practical and industrial use cases.

\subsubsection{RQ 2.3} The aim of this research question is to categorise studies on ABST based on programming languages targeted by these agent-based systems emerging from the papers being studied. Our results suggest that in more than half of the papers (51\%), ABST targets the Java programming language. In almost 37\% of the papers there is no explicit mention of any programming language used. Other programming languages like C, C++, Python, and Perl are rarely targeted.

 \begin{mdframed}[style=style1]
 {\it Answer RQ2: Research in ABST has used reactive, BDI, deliberative and cooperative agent architectures. The majority of these solutions are targeting the Java programming language and are using JADE framework.}
 \end{mdframed}



\subsection{Testing Characteristics}

\subsubsection{RQ 3.1} The aim of this research question is to identify the reported testing level at which ABST approaches are used. The overall results are shown in Figure~\ref{fig:levWels}. Our results suggest that most of the ABST studies (i.e., 28 papers) are targeting system level testing. In addition, 14 of the studies are performing ABST at unit level while just two studies are on integration level testing \footnote{We note here that some ABST approaches focused on multiple testing abstractions levels.}.


\subsubsection{RQ 3.2} The aim of this research question is to categorise studies on ABST based on the testing areas emerging from the papers being studied. As shown in Figure~\ref{fig:levWels}, most of the ABST approaches are targeting functional testing (18 studies), white-box testing (19 studies) and non-functional testing (13 studies). The scarcity of ABST research in regression testing (4 studies) indicates that this area of research has not scaled to regression test selection and continuous integration practices. 


\subsubsection{RQ 3.3} The aim of this research question is to identify the reported types of applications targeted by the ABST approaches. In Figure~\ref{fig:levWels} we show the overall results of this classification. The results suggest that ABST is mostly used in software testing of distributed software applications (46\% of the studies). In addition, ABST targets other application domains such as web-based applications (19.5\% of the studies), network applications (almost 15\% of the studies), and industrial automation systems (19.5\% of the studies). 


 \begin{mdframed}[style=style1]
 {\it Answer RQ3: Most of the ABST approaches are used to perform functional and white-box testing at system level and are targeting web-based and distributed software applications.}
 \end{mdframed}

\section{Related work}
Several mapping studies have been conducted in the software testing research area~\cite{engstrom2015mapping,akbulut2019systematic,hellmann2013agile,lopez2015first} to identify the level of growth, usage, and effects of different types of testing in various application domains. The usage of various technologies for efficient testing has also been studied and mapped. 

Cruz et al.~\cite{cruz2019replication} found that software testing is one of the areas with the most secondary studies and replications in software engineering. Additionally, a mapping study performed by Engström \textit{et al}.~\cite{engstrom2015mapping} identified the potential gap between research and practice in software testing. Several mapping studies have been conducted to study the testing trends in various application domains for testing such as cloud testing~\cite{ahmad2017systematic}, web testing~\cite{akbulut2019systematic}, combinatorial testing~\cite{lopez2015first} and agile testing~\cite{hellmann2013agile}. These studies aim at identifying the gaps in current testing research, as well as trends and directions for future testing research. Recently, test automation and intelligent testing using search-based approaches have been used to improve test efficiency and effectiveness. One of these directions relates to the use of agents in software testing. Norouzi \textit{et al.}~\cite{norouzi2018systematic} conducted a survey to identify the different behaviors, forms and other trends in the use of intelligent agents across various domains like education and virtual assistance. 
To summarize, to the best of our knowledge, this is the first mapping study on agent-based software testing and more studies are needed to improve the study of ABST and its wider application to software development problems. 


\section{Conclusion and Future Work}

The use of agent-based systems in software testing and test automation is a growing area of research that should be recognized and considered.
Therefore, in this study we present a  definition of the
agent-based software testing (ABST) research area and conduct a
systematic mapping study based on several identified concepts and dimensions related to agent-based systems and software testing.

The definition highlights that ABST is the application of agents (e.g., software agents, intelligent agents, autonomous agents, multi-agent systems) to software testing problems by automating complex testing tasks. We found a total of 41 related papers in ABST research area. Our results suggest that there is an interest in this area after 1999 that resulted in the development of solutions using reactive, BDI, deliberative and cooperate agent architectures for software testing. The majority of the implemented ABST approaches are designed using the JADE framework and are using the Java programming language. In addition, our results suggest that the majority of the ABST approaches are used at system-level testing for functional, non-functional and white-box testing. The results of our systematic map also indicate that the current body of knowledge concerning ABST does report only four studies on regression testing. Future work could include the extension of this systematic mapping study into a systematic literature review.

\section*{Acknowledgment}
This work is partially supported by the Swedish Innovation Agency (Vinnova) through the XIVT project, and the UNICORN (Sustainable, Peaceful and Efficient Robotic Refuse Handling) project.
\balance
\bibliographystyle{IEEEtran}
\bibliography{acmart} 

\end{document}